\documentclass{ckm}                 

\usepackage{txfonts}            
\usepackage{epsfig}

\confname{Workshop on the CKM Unitarity Triangle, IPPP Durham, April
  2003}

\title{CP Violation and New Physics\thanks{The analysis
presented in section 3 was done in collaboration with
F.~J.~Botella, M.~Nebot and M.~N.~Rebelo in Ref. \cite{Botella:2002fr}}}

\author{Gustavo C. Branco}
\address{Grupo Teorico de Fisica de Particulas and Departamento de Fisica 
           Instituto Superior Tecnico, Av. Rovisco Pais, P-1049-001 Lisboa, Portugal}

\begin{document}

\begin{abstract}
We describe some of the extensions of the SM, including models with spontaneous CP violation,
where New Physics relevant for CP violation may arise. It is emphasized that the SM predicts 
a series of exact relations among various measurable quantities, such as moduli of CKM
matrix elements and rephasing invariant phases. These exact relations provide a stringent 
test of the SM, with the potential to reveal New Physics.

\end{abstract}

\maketitle


\section{Introduction}
The study of CP violation in its multiple aspects is likely to continue
playing a crucial role in testing the Standard Model(SM) and in searching for New Physics.
So far, all experimental data on flavour physics and CP violation 
\cite{CKMyellow2002}  are in
agreement with the SM \ and its Kobayashi-Maskawa (KM) mechanism \cite{Kobayashi:fv}.
This agreement is impressive,
since one has to accommodate a large number of data with only a few
parameters. The Cabibbo, Kobayashi and Maskawa (CKM) matrix is characterized
by four parameters which one can choose to be the three angles $\theta _{i}$ and
the phase $\delta $ of the standard parametrization \cite{Hagiwara:fs}. 
The values of $s_{1}$, 
$s_{2}$ and $s_{3}$ ( $s_{i}=\sin \theta _{i}$ ) can be determined by the
experimental value of $\left| V_{us}\right| $, $\left| V_{cb}\right| $ 
and $\left| V_{ub}\right| $. Once these parameters are fixed, one has to fit,
using only the phase $\delta $, a large amount of data, 
such as $\varepsilon _{K}$, $\varepsilon ^{\prime }/\varepsilon$, 
$\sin \left( 2\beta \right)$, $\Delta M_{B_{d}}$, 
$\Delta M_{B_{s}}$. It is remarkable that these
five experimental quantities can be fitted with only one 
parameter, namely
the KM phase $\delta $.

In spite of this success of the SM, the search for New
Physics through the study of CP violation phenomena is well motivated by various reasons, 
such as:

i) CP violation is closely related to the least understood aspects of the SM, namely the Higgs
sector and the structure of Yukawa couplings.

ii) Almost any extension of the SM has new sources of CP violation.

iii) CP violation is one of the crucial ingredients needed to generate the Baryon Asymmetry 
of the Universe (BAU). It has been established that the strength of CP violation in the SM 
is not sufficient to generate the observed BAU. Therefore, in all successful baryogenesis scenarios,
including baryogenesis through leptogenesis \cite{Fukugita:1986hr} , 
new sources of CP violation are present.   

iv) Although CP violation can be incorporated in the SM through the introduction of complex 
Yukawa couplings, one would like to have a deeper understanding of the origin of CP violation.
Such an understanding will certainly require a framework of physics beyond the SM.
One may ask, for example, the question whether there are any connections  among the 
various possible manifestations
of CP violation, namely those in the quark sector, and in the leptonic sector. In particular 
one may ask whether there is  
any relation between leptonic CP violation observable in neutrino oscillations and CP violation 
needed for leptogenesis \cite{varios}, 
\cite{Branco:2001pq},  \cite{Rebelo:2002wj}.
Or one may also wonder whether all 
manifestations of CP violation have a common origin \cite{Branco:2003rt}. 
 
In most of the extensions of the SM, it is necessary to control the new sources of CP violation 
in order to conform to the experimental value of  $\varepsilon _K$, 
as well as to the experimental limits on the
electric dipole moments of the neutron and the electron. A notable example is the supersymmetric
extension of the SM, where in general a very large number of new 
phases arise, leading to the so called
supersymmetric CP problem, 
which can be solved by either assuming that the new phases are small
or by having an alternative suppression mechanism \cite{Nir:2001ge}. \\
We will not present a general discussion of models of CP violation 
since it is beyond the scope of this contribution. Instead,
we will divide models of CP violation into two broad classes, 
based on the nature of CP breaking,
which may be spontaneous or explicit. One of the motivations for having 
spontaneous CP violation, 
as emphasized by Lee \cite{Lee:iz} 
in his pioneering work is puting the breaking of CP on the same footing as 
the breaking of gauge symmetry, which is spontaneously broken. Another 
motivation has to do with 
the fact that spontaneous CP breaking provides an alternative solution 
to the strong CP problem \cite{strongcp}, \cite{Nelson}, \cite{Barr:qx} \cite{Bento:ez}
(apart from the Peccei Quinn solution \cite{Peccei:1977ur}) in models where 
$\overline \theta$ naturally vanishes at tree level
and it is calculable in higher orders.
Furthermore, it was shown some time ago that CP can be spontaneously broken 
in string theory \cite{StrWit} 
and more recently it has been pointed out that in some string theory compactifications, CP is an 
exact gauge symmetry and thus its breaking has to be spontaneous \cite{dois}.

\section{ The Breaking of CP: Explicit or Spontaneous? }
Indeed, one of the basic questions one may ask about CP is whether it is explicitly broken at the 
Lagrangian level or, on the contrary, it is a good symmetry of the Lagrangian, spontaneously 
broken by the vacuum. It is remarkable that two of the most interesting models of CP violation,
suggested in the early days of gauge theories were published in the same year (1973) and 
belong to each one of the above categories. We are refering to
the Two-Higgs-Doublet-Model (THDM) suggested by Lee, where CP is 
spontaneously broken and the celebrated KM model \cite{Kobayashi:fv} 
where CP is explicitly
broken at the Lagrangian level, through the introduction of complex Yukawa couplings.
Next we consider some of the simplest extensions of the SM which allow for spontaneous CP violation.

\subsection{The Lee Model}
It can be readily shown that in the SM with only one Higgs doublet, CP cannot be spontaneously
broken. Lee has proposed a minimal extension of the SM where spontaneous CP violation (SCPV)
can be achieved, through the introduction of two Higgs doublets. Due to the presence in the Higgs 
potential of terms of the form $ \Phi^{\dag}_1 \Phi_2
\Phi^{\dag}_1 \Phi_2 $ , $\Phi^{\dag}_1 \Phi_2 \Phi^{\dag}_1
\Phi_1$ , $\Phi^{\dag}_2 \Phi_2 \Phi^{\dag}_1 \Phi_2 $     
the potencial is sensitive to the relative phase between the vevs of the two neutral 
Higgs fields, $ <\phi _j^0> =  v_j \exp (i \theta _j)$.
There is a region of the parameters of the Higgs potential for which its minimum 
corresponds to a non vanishing $\theta = (\theta_2 -
\theta_1)$. 
This leads in general to spontaneous CP violation. At the time 
Lee suggested this model, only two (incomplete) generations were known. In this case, 
the only source of CP violation was Higgs exchange. If one considers the THDM with SCPV 
in the framework of the 3-fermion-generations SM, a non-trivial KM phase is generated in the 
CKM matrix, in spite of the fact that Yukawa couplings are real. This can be readily verified, 
by noting that the quark mass matrices for the down and up quarks have the form:

\begin{equation}
\begin{array}{cc}
M_d &= {1 \over \sqrt{2}} \left[ v_1 e^{i \theta_1} Y^d_1 + v_2 e^{i \theta_2}
Y^d_2 \right] \\
 & \\
M_u &= {1 \over \sqrt{2}} \left[ v_1 e^{- i \theta_1} Y^u_1 + v_2 e^{- i
\theta_2} Y^u_2 \right ] \end{array} 
\end{equation}

where $Y^d_k, Y^u_k$ stand for the Yukawa coupling matrices.
One obtains for the hermitian quark mass matrices:

\begin{eqnarray}
H_d & \equiv M_d M^{\dag}_d  =  {1 \over 2} [ v^2_1 Y^d_1 Y^{dT}_1 + v^2_2
Y^d_2 Y^{dT}_2  \nonumber \\
 & + v_1 v_2 (Y^d_1 Y^{dT}_2 + Y^d_2 Y^{d T}_1) \cos \theta 
 \nonumber\\
 &   -i v_1 v_2 (Y^d_1 Y^{d T}_2 - Y^d_2 Y^{d T}_1) \sin \theta]  
\label{hdd} \\
H_u &\equiv M_u M^{\dag}_u  =  {1 \over 2} [ v^2_1 Y^u_1 Y^{uT}_1 + v^2_2
Y^u_2 Y^{uT}_2  \nonumber \\
 & + v_1 v_2 (Y^u_1 Y^{uT}_2 + Y^u_2 Y^{u T}_1) \cos \theta 
\nonumber\\ 
 &  +i v_1 v_2 (Y^u_1 Y^{u T}_2 - Y^u_2 Y^{u T}_1) \sin \theta ]
\label{huu} 
\end{eqnarray}

Although there is only one physical phase, namely $\theta = (\theta_2 -
\theta_1)$, it is clear from Eqs. (\ref{hdd}) and (\ref{huu}) that due to 
the arbitrariness of
$Y^d_k, Y^u_k$, the matrices $H_u, H_d$ are arbitrary hermitian matrices. As
a result, there will be in general a non-trivial CP violating phase in the
CKM matrix. For any specific choice of $Y^d_k, Y^u_k$ this can be explicitly
verified by computing the invariant \cite{Bernabeu:fc} quantity
$T \equiv  {\mbox tr} [H_u, H_d]^3 $. 
For three fermion generations the non-vanishing of this weak-basis invariant 
is a necessary and sufficient condition for having CP violation
mediated by charged weak-interactions.

   From the above discussion, one concludes that in the three fermion 
generations version of the Lee model, one has two sources of CP violation, 
namely:

i) The usual KM mechanism contributing to CP violation in decay amplitudes 
as well as to $K^{0}-\overline K^{0}$,
$B_{d}^{0}-\overline{B}_{d}^{0}$  and 
$B_{s}^{0}-\overline{B}_{s}^{0}$ 
mixings through the usual box diagrams.

ii) Flavour-changing neutral Higgs mediated interactions giving additional
tree level contributions to the neutral meson mixings mentioned in i).

The above generalization of the Lee model to three fermion generations illustrates
a very common situation, where one has the KM mechanism, together with other
sources of CP violation.

One of the potential drawbacks of the Lee model, is the fact that 
the existence of Higgs mediated flavour-changing neutral currents(FCNC) 
at tree level requires very heavy neutral scalars, of the order of a few Tev, 
unless there is some suppression mechanism \cite{aaaaa}. 
It has been shown that in the framework 
of two-Higgs-doublets models, the introduction of appropriate 
discrete symmetries \cite{Branco:1996bq}
leads to the suppression of the FCNC vertex between, for example, two down-type
quarks i and j by products of the CKM matrix elements 
of the type  $V_{\alpha i}^\ast V_{\alpha j}$,
where $\alpha$ denotes one of the up-type quarks.
In the case $\alpha = t$, $i=d$, and $j=s$ this suppression is quite strong,
and in that class of models neutral Higgs may be relatively light
(e.g. 100-200 GeV), even in the presence of Higgs mediated FCNC.

\subsection{Multi-Higgs Models with Natural Flavour Conservation}

One may, of course, eliminate altogether FCNC in the two-Higgs-doublets
models by implementing Natural Flavour Conservation (NFC) in the Higgs sector
through a $Z_2$ discrete symmetry \cite{Glashow:1976nt}. 
However, in this case, the structure 
of the Higgs potential is such that no spontaneous CP violation can be achieved,
unless the discrete symmetry is softly broken \cite{Branco:1985aq}.
If one insists on NFC, then a minimum number of three Higgs doublets are necessary
in order to achieve spontaneous CP violation \cite{Branco:1980sz}. 
This class of models with 
NFC, three Higgs doublets and SCPV has the special feature
that the CKM matrix is real \cite{Branco:1979pv}
and CP violation arises exclusively through 
Higgs exchange \cite{Branco:1985pf}. 
The essential reason why the CKM is real in this case
has to do with the fact that the $Z_2$ symmetry constrains $d_R$ to couple to 
only one of the Higgs doublets ( and similarly for $u_R$). In this case,
any phase can be rotated away from the quark mass matrices, through
a redefinition of the righthanded quark fields.
In the version of the three Higgs-doublet model, with
explicit CP violation \cite{Weinberg:1976hu}
and in the presence of three fermion generations, 
one has again two sources of CP violation, namely the KM mechanism and 
Higgs exchange.

\subsection{SCPV in Supersymmetric Extensions of the SM}

The Minimal Supersymmetric Standard Model (MSSM) has two Higgs doublets and
therefore it is a natural candidate to achieve SCPV. However, it is not possible to obtain
SCPV in the MSSM at tree level, due essentially to the fact that SUSY does not allow
some of the couplings which are present in the general THDM.
Since SUSY has to be softly broken, radiative corrections can induce new CP
violating operators which could induce CP breaking \cite{Maekawa}. 
However, the possibility 
that radiative corrections can cause SCPV, requires the existence of a light 
scalar \cite{vvv}
which is excluded by LEP. It has been shown that one may achieve SCPV in the 
Next-to-Minimal-Supersymmetric-Standard-Model (NMSSM) \cite{Pomarol2},
\cite{Glasgow}, \cite{Lisbon}
where a singlet superfield
is added to the Higgs sector. In this case, the CKM matrix 
is real \cite{Branco:1979pv}, essentially 
due to the same reason explained above for the 3-Higgs doublet model with SCPV
and NFC. In the NMSSM with SCPV all couplings are real, so that CP is a good symmetry 
of the Lagrangian. However the physical relative phases of the Higgs doublet and 
singlet enter in the chargino and neutralino mass matrices as well as in some 
vertices. As a result, it has been shown that chargino box diagrams 
can generate \cite{Lisbon} the 
observed experimetal value of $\varepsilon _K$.
As far as  $\varepsilon ^{\prime }/\varepsilon$, and  $a_ {J/\psi K_s}$,
it has been pointed out that SUSY contributions in these models can saturate the 
experimental values \cite{Pomarol2} \cite{Lebedev}, 
provided there is maximal LR squark mixing. 

\subsection{Spontaneous CP Violation Generated at a High Energy Scale}
If one maintains the fermion spectrum of the SM, the THDM suggested by
Lee has the simplest Higgs structure needed to generate spontaneous CP breaking
capable of accounting for the experimentally observed CP violation. However,
it is possible to generate relevant spontaneous CP violation with only one Higgs 
doublet $\phi$ and one complex scalar singlet $S$, provided that one also introduces at least 
one  singlet charge $- \frac{1}{3}$ vectorial quark $D^0$. 
The scalar potential will contain terms in $\phi$
and S with no phase dependence, together with terms of 
the form \mbox{$({\mu }^2 + \lambda_1
 S^\ast S +\lambda_2 {\phi ^ \dagger } \phi )(S^2 + S^{\ast 2}) +
\lambda_3 (S^4 + S^{\ast 4})$} which, in general, lead to 
the spontaneous breaking of T and CP invariance  \cite{Bento:1990wv}
with $\phi$ and $S$ acquiring vacuum expectation values (vevs) 
of the form:
\begin{equation}
\langle {\phi}^0 \rangle = \frac{v}{\sqrt 2}, \ \   \ \ \   
\langle S \rangle = \frac{V \exp (i \alpha )}{\sqrt 2}
\label{vev}
\end{equation}

In this class of models the presence of
the vector-like quark $D^0$ plays a crucial r\^ ole, since 
it is through the couplings 
$({f_q} S + {f_q}^{\prime} S^\ast ) {\overline {D_L^0}} d_R^0$
that the phase $\alpha$ appears in the effective mass matrix 
for the down standard-like quarks. It can be shown that the phase
$\delta _{KM}$, generated through spontaneous CP violation
is not suppressed by factors of 
$\frac{v}{V}$ . 
For very large $V$ (e.g. $V$ 
$\sim M_{GUT}\sim 10^{15} $ Gev), $\delta _{KM}$ is the only leftover 
effect at low energies,
from spontaneous CP breaking at high energies. For not so
large a value of $V$ (e.g., $V$ of the order of a few Tev) the
appearance of significant flavour changing neutral currents
(FCNC) in the down quark sector leads to new contributions to
$B_d - \overline{B_d} $ and $B_s - \overline{B_s} $ 
mixing which can alter  \cite{fcnc} some of the predictions of the
SM for CP asymmetries in B meson decays. These FCNC are
closely related to the non-unitarity of the
$3 \times 3$ CKM matrix, with both effects suppressed by powers of 
$\frac{v}{V}$.\\
This class of models has been extended to the leptonic sector 
where the role of vector-like quarks is played by the righthanded
neutrinos. It was pointed out \cite{Branco:2003rt} that in such 
a framework, all 
manifestations of CP violation may have a common origin. In particular,
the phase $\alpha$ defined in Eq.(4) generates CP violation in the 
quark sector, in the leptonic sector at low energies (measurable 
for example in neutrino oscillations), as well as CP violation 
required by leptogenesis.

\section{Precision Tests of the SM and the Search for New Physics}
From the discussion in the previous section it should be clear that
when one considers extensions of the SM, the most common situation is
having the usual KM mechanism, together with other sources of CP violation.
This is of course the case when one assumes that CP is explicitly broken 
at the Lagrangian level. What is remarkable, is the fact that it is
also true for some of the models of SCPV, where all the couplings 
of the Lagrangian are real.\\
In our analysis, we will assume that the tree level weak decays
are dominated by the SM W-exchange diagrams, thus implying that the 
extraction of $\left| V_{us}\right| $, $\left| V_{ub}\right| $ 
and $\left| V_{cb}\right| $ from experiment continues to be valid
even in the presence of New Physics (NP). We will allow for 
contributions from NP in processes like 
$B_{d}^{0}-\overline{B}_{d}^{0}$ mixing and 
$B_{s}^{0}-\overline{B}_{s}^{0}$ mixing, as well as in penguin diagrams.
Since the SM contributes to these processes only at loop level, the
effects of NP are more likely to be detectable. Examples of processes
which are sensitive to NP, are the CP asymmetries corresponding to the
decays ${B^0}_d \rightarrow J/\Psi K_{s}$ and
${B^0}_d \rightarrow \pi^+ \pi^-$ which are affected by NP contributions
to $B_{d}^{0}-\overline{B}_{d}^{0}$ mixing. Significant contributions
to $B_{d}^{0}-\overline{B}_{d}^{0}$ and 
$B_{s}^{0}-\overline{B}_{s}^{0}$ mixing can arise in many of
the extensions of the SM, such as models with vector-like quarks
\cite{fcnc} and supersymmetric extensions 
of the SM \cite{susy}. Vector-like quarks naturally 
arise in theories with large extra-dimensions \cite{extra},
as well as in some grand-unified theories like $E_6$. 
As previously mentioned, the presence of vector-like 
quarks leads to a small deviation of
$3 \times 3$ unitarity of $V_{CKM}$ which in turn leads to Z-mediated 
new contributions to $B_{d}^{0}-\overline{B}_{d}^{0}$ and 
$B_{s}^{0}-\overline{B}_{s}^{0}$ mixings. In the minimal 
Supersymmetric Standard Model (MSSM) the size of SUSY
contributions to $B_{d}^{0}-\overline{B}_{d}^{0}$  and 
$B_{s}^{0}-\overline{B}_{s}^{0}$ mixing crucially depends
on the choice of soft-breaking terms, but there 
is a wide range of the parameter space where SUSY
contributions can be significant. Recently, it has been
pointed out \cite{Chang:2002mq} that in the context of SUSY SO(10),
there is an interesting connection between the observed large 
mixing in atmospheric neutrinos and the size of the SUSY
contribution to  $B_{s}^{0}-\overline{B}_{s}^{0}$ mixing,
which is expected to be large in this class of models.\\
The standard way of testing the compatibility of the SM with
the existing data consists of adopting the 
Wolfenstein parametrization \cite{Wolfenstein:1983yz} and
plotting in the $\rho $, $\eta $ plane the constraints derived from various
experimental inputs, like the value of $\varepsilon _{K}$, the size of 
$\left| V_{ub}\right| /\left| V_{cb}\right| $,
the value of $a_ {J/\psi K_s}$, as well as the strength of 
$B_{d}^{0}-\overline{B}_{d}^{0}$ 
and $B_{s}^{0}-\overline{B}_{s}^{0}$ mixings. The challenge 
for the SM is then to find a region in 
the $\rho $, $\eta $ plane where all the constraints are simultaneously 
satisfied.
A complementary way of testing the SM, consists of
using exact relations connecting measurable
quantities, namely moduli of $V_{CKM}$ and the arguments of 
rephasing invariant quartets. These relations can be derived 
in the framework of the SM and follow from the implicit
assumption of unitarity of $V_{CKM}$. They have the interesting
feature of being independent of any particular parametrization
of the quark mixing matrix.

\subsection{Choice of Rephasing Invariant Phases}
Using the freedom to rephase quark fields, it can be readily shown 
that the $3 \times 3$ sector of a CKM matrix of arbitrary size 
contains only four independent rephasing invariant phases. It is
convenient to make the following choice:
\begin{equation}
\begin{array}{c}
\gamma \equiv \arg (-V_{ud}V_{cb}V_{ub}^{\ast }V_{cd}^{\ast })=\arg 
\left( -\frac{V_{ud}V_{ub}^{\ast }}{V_{cd}V_{cb}^{\ast }}\right)  \\ 
\beta \equiv \arg (-V_{cd}V_{tb}V_{cb}^{\ast }V_{td}^{\ast })=\arg
\left( -\frac{V_{cd}V_{cb}^{\ast }}{V_{td}V_{tb}^{\ast }}\right)  \\ 
\chi \equiv \arg (-V_{cb}V_{ts}V_{cs}^{\ast }V_{tb}^{\ast })=\arg 
\left( -\frac{V_{cb}V_{cs}^{\ast }}{V_{tb}V_{ts}^{\ast }}\right)  \\ 
\chi ^{\prime }\equiv \arg (-V_{us}V_{cd}V_{ud}^{\ast }V_{cs}^{\ast })
=\arg \left( -\frac{V_{us}V_{ud}^{\ast }}{V_{cs}V_{cd}^{\ast }}\right) 
\end{array}
\label{fases}
\end{equation}
Furthermore, in order to fix the invariant phases entering in
$B^{0}$ CP asymmetries, it
is useful to adopt the following phase convention~\cite{Branco:1999fs}:
\begin{equation}
\arg (V)=\left( 
\begin{array}{lll}
0 & {\chi ^{\prime }} & -\gamma  \\ 
\pi  & 0 & 0 \\ 
-\beta  & \pi +\chi  & 0
\end{array}
\right) \label{fasesconv}
\end{equation}
Through the measurement of CP asymmetries, one can obtain the phases of the
rephasing invariant quantities:
\begin{eqnarray}
\lambda _{f}^{\left( q\right) }=\left( \frac{q_{B_{q}}}{p_{B_{q}}}\right)
\left( \frac{A\left( \overline{B}_{q}^{0}\rightarrow f\right) }{A\left(
B_{q}^{0}\rightarrow f\right) }\right) ; \nonumber \\  
\lambda _{\overline{f}}^{\left( q\right) }=
\left( \frac{q_{B_{q}}}{p_{B_{q}}}\right) \left( 
\frac{A\left( \overline{B}_{q}^{0}\rightarrow \overline{f}\right) }
{A\left(B_{q}^{0}\rightarrow \overline{f}\right) }\right) \label{lambdas}
\end{eqnarray}
The first factor in $\lambda _{f}^{\left( q\right) }$ is due to mixing and
its phase equals $(-2\beta )$ and $2\chi $ for $B_{d}$ and $B_{s}$,
respectively. Let us consider the general case where New Physics
(NP) also contributes to the mixing. It is convenient to parametrize
the NP contributions in the following way:
\begin{eqnarray}
M_{12}^{\left( q\right) } =\left( M_{12}^{\left( q\right) }\right)
^{SM}r_{q}^{2}e^{-2i\phi _{q}}\Rightarrow \nonumber \\
\Delta M_{B_{q}}=\left( \Delta
M_{B_{q}}\right) ^{SM}r_{q}^{2} \label{Munodos}\\
  \nonumber \\
\frac{q_{B_{q}}}{p_{B_{q}}} =\exp \left( i\arg \left( M_{12}^{\left(
q\right) }\right) ^{\ast }\right) =\left( \frac{q_{B_{q}}}{p_{B_{q}}}\right)
^{SM}e^{2i\phi _{q}}\label{mixing}
\end{eqnarray}
In the presence of NP, the phases from mixing become 
$2(-\beta +\phi_{d})\equiv - 2\overline{\beta }$ and 
$2(\chi +\phi _{s})\equiv 2\overline{\chi }$ for 
$B_{d}$ and $B_{s}$ decays, respectively. It is clear that $r_{q}\neq 1$ 
and/or $\phi _{q}\neq 0$ would signal the presence of NP.
It is not easy to separate $\beta$ from a possible NP 
contribution ($\phi_{d}$) in $B_{d}^0$ decays
like ${B^{0}}_{d}\rightarrow J/\Psi K_{s}$. This
renders specially important the measurement of $\gamma $, which does not
suffer from contamination of NP in the mixing. Note that $\gamma $ can be
either directly measured \cite{gamma} 
or obtained through the knowledge of the
asymmetries $a_{J/\Psi K_{s}}=\mbox{Im} \left( 
\lambda _{J/\Psi K_{s}}^{\left( d\right) }\right) $, 
$a_{\pi ^{+}\pi ^{-}}=\mbox{Im} \left(
\lambda _{\pi ^{+}\pi ^{-}}^{\left( q\right) }\right) $. 
Indeed the phase $\phi _{d}$ cancels in the sum 
$\overline{\alpha }+\overline{\beta }=\left(
\pi -\gamma -\beta +\phi _{d}\right) +\left( \beta -\phi _{d}\right) $ and
one has:
\begin{equation}
\gamma =\pi -\frac{1}{2}\left[ \arcsin a_{J/\Psi K_{s}}+\arcsin a_{\pi
^{+}\pi ^{-}}\right]\label{arcsin} 
\end{equation}
Note that we are using $a_{\pi^{+}\pi ^{-}} = \sin (2 \overline{\alpha})$
that can be extracted from the experimental asymmetry through
various different approaches \cite{app}.
Once $\gamma $ is known, $\beta $ can be readily obtained, using unitarity
and the knowledge of $\left| V_{ub}\right| $, $\left| V_{us}\right| $, $\left| V_{cb}\right| $. 
The knowledge of $\beta $, together with $a_{J/\Psi
K_{s}}$ leads then to the determination of $\phi _{d}$. Of course, this
evaluation of $\phi _{d}$ will be restricted by the precision on $\left|
V_{ub}\right| $, since $\left| V_{us}\right| $, $\left| V_{cb}\right| $ are
extracted from experiment with good accuracy. Similar considerations apply to
the extraction of $r_{d}$, $r_{s}$ or $r_{d}/r_{s}$ from $\Delta M_{B_{d}}$
and $\Delta M_{B_{s}}$ where $\left| V_{td}^{\ast }V_{tb}\right| $, 
$\left| V_{ts}^{\ast }V_{tb}\right| $ or its ratio, 
have to be reconstructed previously using unitarity.

\subsection{Exact Relations}
Using orthogonality of different rows and different columns of $V_{CKM}$,
one can obtain various exact relations involving moduli and rephasing
invariant phases, such as:
\begin{eqnarray}
\sin \chi =\frac{\left| V_{td}\right| \left| V_{cd}\right| }{\left|
V_{ts}\right| \left| V_{cs}\right| }\sin \beta  \label{rel5} \\
\left| V_{ub}\right| =\frac{\left| V_{cd}\right| \left| V_{cb}\right| 
}{\left| V_{ud}\right| }\frac{\sin \beta }{\sin (\gamma +\beta )}
\label{rel9} \\
\sin \chi =\frac{\left| V_{us}\right| \left| V_{cd}\right| \left|
V_{cb}\right| }{\left| V_{ts}\right| \left| V_{tb}\right| \left|
V_{ud}\right| }\frac{\sin \beta \sin (\gamma +\chi ^{\prime })}{\sin (\gamma
+\beta )}  \label{rel15} \\
\frac{\sin \chi }{\sin (\gamma +\chi ^{\prime })}=\frac{\left|
V_{us}\right| \left| V_{ub}\right| }{\left| V_{ts}\right| \left|
V_{tb}\right| }  \label{rel8} \\
\left| V_{td}\right| =\frac{\left| V_{cd}\right| \left| V_{cb}\right| 
}{\left| V_{tb}\right| }\frac{\sin \gamma }{\sin (\gamma +\beta )}
\label{rel10} 
\end{eqnarray}
Since the above formulae have the potential of providing precise tests of
the SM, we have opted for writing exact relations. However, it is obvious
that given the experimental knowledge on the size of the various moduli of
the CKM matrix elements, some of the above relations can be, to an excellent
approximation, substituted by simpler ones. For example, Eq.(\ref{rel15}) is
the exact version of the Aleksan-London-Kayser relation \cite{Aleksan:1994if}, 
the importance of which has been 
emphasized by Silva and Wolfenstein \cite{Silva:1996ih} : 
\begin{equation}
\sin \chi \simeq \frac{\left| V_{us}\right| ^{2}}{\left| V_{ud}\right| ^{2}}
\frac{\sin \beta \sin \gamma }{\sin (\gamma +\beta )}  \label{rel16}
\end{equation}
Similarly Eq.(\ref{rel5}) can be well approximated by:
\begin{equation}
\sin \chi \simeq r \frac {\left| V_{us}\right| } {\left| V_{ud}\right| }
\sin \beta  \label{rel17}
\end{equation}
while Eqs.(\ref{rel10}) and (\ref{rel8}) lead, respectively to: 
\begin{eqnarray}
r &\simeq &\left| V_{us}\right| \frac{\sin \gamma }{\sin (\gamma +\beta )}
\label{rel18} \\
\sin \chi &\simeq &\frac{\left| V_{us}\right| \left| V_{ub}\right| }{\left|
V_{cb}\right| }\sin \gamma  \label{rel19}
\end{eqnarray}
It is worthwhile to illustrate how these relations can be used to test 
the SM:
\begin{itemize}
\item[(i)]  Eq.(\ref{rel5}) and its approximate form Eq.(\ref{rel17}) would
provide an excellent test of the SM, once $\chi $, $r$ and $\beta $ are
measured. Note that the theoretical errors in extracting $r\equiv \left|
V_{td}\right| /\left| V_{ts}\right| $ from $B_{d}^{0}-\overline{B}_{d}^{0}$
and $B_{s}^{0}-\overline{B}_{s}^{0}$ mixings are much smaller than those
present in the extraction of $\left| V_{td}\right| $, $\left| V_{ts}\right| $.
\item[(ii)]  Eq.(\ref{rel19}) has the important feature of only involving
quantities which are not sensitive to the possible presence of New Physics
in $B_{d}^{0}-\overline{B}_{d}^{0}$ mixing. It has, of course, the
disadvantage of requiring the knowledge of $\left| V_{ub}\right| $ with
significant precision, in order to be a precise test of the SM.
\item[(iii)]  Eq.(\ref{rel18}) gives, to an excellent approximation, 
$r$ in terms of $\gamma $ and $\beta $. This
relation will provide an important test of the SM once $r$, $\gamma $ and $
\beta $ are measured. Note that in the SM, one knows that $r$ is of order 
$\left| V_{us}\right| $, the importance of Eq.(\ref{rel18}) is that it
provides the constant of proportionality.
\end{itemize}
In the context of the SM the above formulae can 
also be very useful for a precise
determination of $V_{CKM}$ from input data: for example, if $\beta $ and 
$\gamma $ are measured with sufficient accuracy, one can use 
Eqs.(\ref{rel9}), (\ref{rel10}) to determine $\left| V_{ub}\right| $, 
$\left| V_{td}\right| $. 
One can thus reconstruct the full CKM matrix, using $\left| V_{us}\right| 
$, $\left| V_{cb}\right| $, $\beta $ and $\gamma $ as input parameters.
Furthermore we can also predict the SM value for $\sin 2\chi $ and $\sin
\chi ^{\prime }$.

The above expressions can also be used to detect, in a quantitative way,
the presence of New Physics in $B_{d}^{0}-\overline{B}_{d}^{0}$ and or 
$B_{s}^{0}-\overline{B}_{s}^{0}$ mixings. For example,
Eq.(\ref{rel9}), we see that this unitarity relation
can only be affected by the presence of $\phi _{d}$, therefore this
equation allows for a clean extraction of $\phi _{d}$.
By writing Eq.(\ref{rel9}) in terms of $\overline{\beta }$ 
and $\phi _{d}$ (note that $Im\left( \lambda _{J/\Psi K_{s}}^{(d)}\right)
 =\sin \left( 2\overline{\beta } \right) $) we get 
\begin{equation}
\tan \left( \phi _{d}\right) =\frac{R_{u}\sin \left( \gamma +\overline{\beta 
}\right) -\sin \left( \overline{\beta }\right) }{\cos \left( \overline{\beta 
}\right) -R_{u}\cos \left( \gamma +\overline{\beta }\right) }  \label{phid}
\end{equation}
with 
\begin{equation}
R_{u}=\frac{\left| V_{ud}\right| \left| V_{ub}\right| }{\left| V_{cd}\right|
\left| V_{cb}\right| }  \label{Ru}
\end{equation}

\begin{figure}[h]
\begin{center}
\epsfig{file=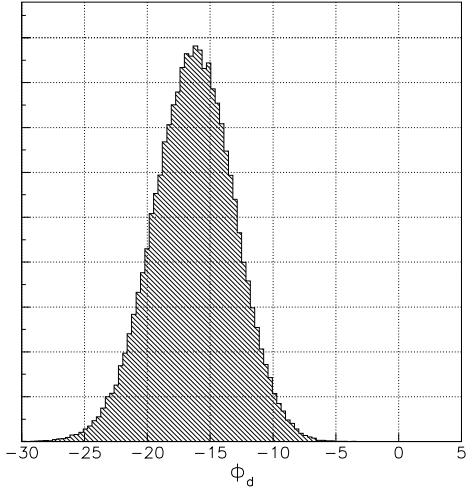,width=2.8781in}
\end{center}
\caption{The $\protect\phi _{d}$
distribution in degrees corresponding to an example where New Physics
is clearly detected.}
\label{fidnpcaso1}
\end{figure}

From Eq.(\ref{phid}), we can find out the bounds that can be reached for 
$\phi _{d}$, once we have a direct measurement of $\gamma $. 
To illustrate the usefulness of Eqs.(\ref{rel9}) and (\ref{phid})
one can consider 
examples of different sets of assumed data  which hopefully will be 
available in the near future. For definiteness let us consider the 
most optimistic scenario, where NP is discovered, corresponding to
the following example \footnote{For more examples see 
Ref~\cite{Botella:2002fr}}:
\begin{eqnarray}
\left| V_{us}\right| =0.221\pm 0.002  \nonumber \\
\left| V_{cb}\right| =0.0417\pm 0.0010 \nonumber \\
\left| V_{ub}\right| =\left( 4.05\pm 0.21\right) \times 10^{-3} \\ 
{\overline \beta} =\left( 30.0\pm 0.3\right){{}^o}
 \gamma =\left( 20 \pm 5\right){{}^o} \nonumber
\end{eqnarray}
the resulting $\phi_{d}$ distribution is presented in 
Fig.\ref{fidnpcaso1} corresponding
to $\phi_{d}=\left( -16.3 \pm 3.2\right){{}^o}$
In this case, one would have a clear indication 
of NP in the phase of $B_{d}^{0}-\overline{B}_{d}^{0}$ mixing.
Note that for this choice of $\gamma$ the value of 
$\varepsilon _{K}$ would not be saturated by the SM
contribution. Therefore, in this example one would conclude 
that NP also contributes to 
$\varepsilon _{K}$.

\section{Conclusions}
We have described the main features of CP violation in a variety
of models beyond the SM, emphasizing that the most common situation is 
having the KM mechanism together with some extra sources of CP violation. 
Often this New Physics gives additional contributions to 
$B_{d}^{0}-\overline{B}_{d}^{0}$ and $B_{s}^{0}-\overline{B}_{s}^{0}$ mixings
which can affect the predictions for the various CP asymmetries.
Furthermore, we have pointed out that the SM predicts a series of exact relations
connecting measurable quantities like 
moduli and rephasing invariant quartets of the CKM matrix
which provide a stringent test of the SM, with the potential of
revealing New Physics. This is specially true if,
on the one hand, $\gamma$, $x_s$ and eventually $\chi$ are
measured in the present or future B factories and, on the other 
hand, there is a significant decrease in the theoretical
uncertainties in the evaluation of the relevant hadronic
matrix elements. These tests may complement the standard 
analysis in the $\rho$, $\eta$ plane.

In the search for New Physics through CP violation,
the first step is, of course, to find a clear deviation 
from the predictions of the SM for flavour physics and CP violation. 
If the need for New Physics is
established through, for example, the appearance of 
new contributions to $B_{d}^{0}-\overline{B}_{d}^{0}$ and/or
$B_{s}^{0}-\overline{B}_{s}^{0}$ mixings, a much more difficult task will 
be to differentiate among the various models where such new contributions
may arise.

\section*{Acknowledgments}
We thank the organizers and convenors for warm hospitality and
for a stimulating Workshop. The participation at CKM03 was
partially supported by a NATO Collaborative Linkage Grant. The work
presented received partial support from Funda\c c\~ ao para
a Ci\^ encia e a Tecnologia (Portugal) through projects 
CFIF-Plurianual (2/91),
CERN/P/FIS/40134/2000,
CERN/FIS/43793/2001, CERN/C/FIS/40139/2000
and from the EC under contract HPRN-CT-2000-00148.

\end{document}